\documentclass[a4paper,fleqn,usenatbib]{mnras}

\usepackage{times}

\usepackage[T1]{fontenc}
\usepackage{ae,aecompl}

\usepackage{savesym}
\usepackage{amsmath}
\savesymbol{iint}
\usepackage{txfonts}
\restoresymbol{TXF}{iint}

\usepackage{graphicx}	
\usepackage{amssymb}	

\newcommand{\numax}{\mbox{$\nu_{\rm max}$}}

\newcommand{\muHz}{\mbox{$\mu$Hz}}
\newcommand{\teff}{\mbox{$T_{\rm eff}$}}

\newcommand{\msun}{\mbox{$\mathrm{M}_{\sun}$}}
\newcommand{\rsun}{\mbox{$\mathrm{R}_{\sun}$}}

\title[Bright Star \emph{Kepler} Smear Photometry]{Photometry of Very Bright Stars with \emph{Kepler} and K2 Smear Data}

\author[B. J. S. Pope et al.]{B. J. S. Pope,$^{1}$\thanks{E-mail: benjamin.pope@physics.ox.ac.uk}
T. R. White$^{2,3}$,
D. Huber$^{4,5,6}$,
S. J. Murphy$^{4,6}$,
T. R. Bedding$^{4,6}$,
\newauthor D. A. Caldwell$^{7,5}$,
A. Sarai$^{4}$,
S. Aigrain$^{1}$,
 and T. Barclay$^{7,8}$
\\
$^{1}$Oxford Astrophysics, University of Oxford, Keble Rd, Oxford OX1 3RH, UK\\
$^{2}$Institut f\"{u}r Astrophysik, Georg-August-Universit\"{a}t G\"{o}ttingen, Friedrich-Hund-Platz 1, 37077 G\"{o}ttingen, Germany\\
$^{3}$Max-Planck-Institut f\"ur Sonnensystemforschung, Justus-von-Liebig-Weg 3, 37077 G\"ottingen, Germany\\
$^{4}$Sydney Institute for Astronomy, School of Physics, University of Sydney, NSW 2006, Australia\\
$^{5}$SETI Institute, 189 Bernardo Avenue, Suite 100, Mountain View, CA 94043, USA\\
$^{6}$Stellar Astrophysics Centre, Department of Physics and Astronomy, Aarhus University, Ny Munkegade 120, DK-8000 Aarhus C, Denmark\\
$^{7}$NASA Ames Research Center, M/S 244-40, Moffett Field, CA 94035, USA\\
$^{8}$Bay Area Environmental Research Institute, Inc., 560 Third St West, Sonoma, CA 95476, USA
}

\date{Accepted XXX. Received YYY; in original form ZZZ}

\pubyear{2015}

\begin{document}
\label{firstpage}
\pagerange{\pageref{firstpage}--\pageref{lastpage}}
\maketitle

\begin{abstract}
High-precision time series photometry with the \emph{Kepler} satellite has been crucial to our understanding both of exoplanets, and via asteroseismology, of stellar physics. After the failure of two reaction wheels, the \emph{Kepler} satellite has been repurposed as \emph{Kepler}-2 (K2), observing fields close to the ecliptic plane. As these fields contain many more bright stars than the original \emph{Kepler} field, K2 provides an unprecedented opportunity to study nearby objects amenable to detailed follow-up with ground-based instruments. Due to bandwidth constraints, only a small fraction of pixels can be downloaded, with the result that most bright stars which saturate the detector are not observed. We show that engineering data acquired for photometric calibration, consisting of collateral `smear' measurements, can be used to reconstruct light curves for bright targets not otherwise observable with \emph{Kepler}/K2. Here we present some examples from \emph{Kepler}~Quarter~6 and K2~Campaign~3, including the $\delta$~Scuti variables HD~178875 and 70~Aqr, and the red giant HR~8500 displaying solar-like oscillations. We compare aperture and smear photometry where possible, and also study targets not previously observed. These encouraging results suggest this new method can be applied to most \emph{Kepler} and K2 fields.
\end{abstract}




\section{Introduction}
\label{intro}

The results of the \emph{Kepler} Mission, launched in~2009, have revolutionized both exoplanetary science \citep{doi:10.1146/annurev-astro-082214-122246} and asteroseismology \citep{2013ARA&A..51..353C}. 
The original \emph{Kepler} Mission ended after four years, following the failure of two reaction wheels required to stabilize the spacecraft pointing.  The K2 Mission is now underway, using the two remaining wheels to control two axes while the third, the overall roll angle, sits at an equilibrium balanced by solar radiation pressure \citep{2014PASP..126..398H}. This requires that the telescope be pointed approximately orthogonal to the direction of the Sun, so that fields are limited to $\sim 80$~day `Campaigns' (hereafter~C) centred close to the ecliptic plane. The first results from asteroseismology with K2 are very promising, detecting solar-like oscillations with observations of red giants subgiants and classical pulsators \citep{2015ApJ...809L...3S,2015arXiv150701827C,2015ApJ...806...30L}.

Because of limitations on communications bandwidth, not all targets `on silicon' (i.e. in the spacecraft field of view) are downloaded. Instead, `postage~stamps' of active pixels are downloaded around predetermined targets of interest, and data from other pixels are discarded. The situation is particularly difficult for the brightest stars: when these sources saturate the detector, electrons bleed along the saturated columns, with the result that very large apertures are required to capture all of the stellar flux. Therefore these brightest stars are very expensive in terms of the limited number of pixels that can be downloaded, and many were omitted from the target lists. The situation is even more difficult in K2 because the `postage stamps' must be larger in order to account for the apparent motion of the sources as the roll angle varies, and also because the ecliptic target fields contain a much higher density of nearby bright stars.

Beginning with the K2 Campaign 3 data release, engineering data were made publicly available on the Mikulski Archive for Space Telescopes (MAST). These included collateral data for both long and short cadence, which consist of measurements of photometric `smear' for each column in the CCD. Since the \emph{Kepler} camera lacks a shutter, light falling on the detector during the read-out stage causes photometric smear that must be calibrated. The short cadences (SC) consist of the sum of nine 6.02~s exposures for a total integration of 54.2~s, and the long cadences (LC) are the sum of 30 times these, for a total of 1626~s. The light landing during read-out is added column-by-column and stored as `collateral' smear data at each cadence of both the \emph{Kepler} and K2 Missions. \citet{orig_smear} suggested using these smear pixels data to obtain photometry of otherwise unobserved \emph{Kepler} stars. They outlined a method for extracting light curves from these data and presented several examples, but to our knowledge this has not been followed up in subsequent work. The goal of this Letter is to demonstrate that smear pixels can be used to extract scientifically useful photometry for very bright stars in both \emph{Kepler} and K2 fields.

The upcoming Transiting Exoplanet Survey Satellite (TESS) \citep{2015JATIS...1a4003R} Mission aims to study such nearby, bright stars to identify promising candidates for further study, but many K2 targets will be out of its reach. TESS will observe polar regions most frequently and devote significantly less time to equatorial fields, with the result that the ecliptic objects studied by K2 are excluded from most of the nominal TESS Mission. Therefore, for these targets, K2 smear data provide the only space-based photometry likely to be available in the near future.

\section{Method}
\label{method}

\begin{figure*}
\begin{center}
	\includegraphics[width=0.8\textwidth]{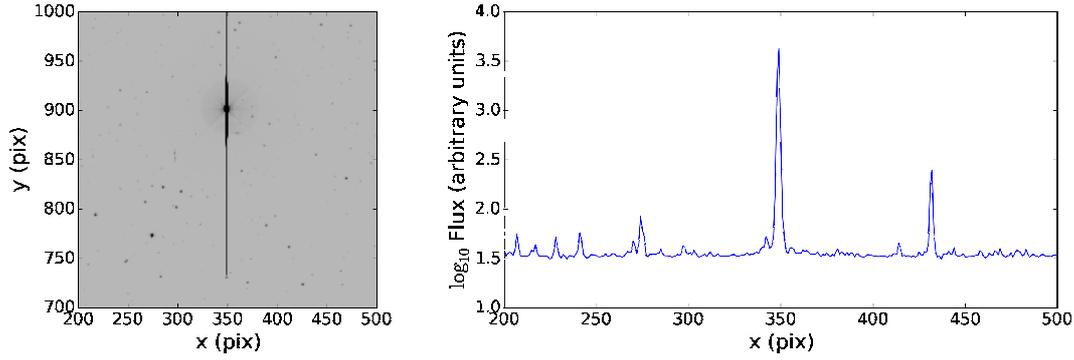}
	\caption{Smear data from K2~Campaign~3 for the bright star HR~8500.
    Left: Part of the full-frame image, including HR~8500 as the saturated star with a long bleed column extending above and below. Right: 1D masked smear profile, with HR 8500 clearly visible as the highest peak.
	}
	\label{fig:HR8500_ffi}
\end{center}
\end{figure*}

The smear data consist of FITS tables containing `virtual' and `masked' smear flux. These contain, respectively, 
smear information from 12 masked rows at the start of each read-out and 12 over-clocked rows after each frame is read out. The 12 smear rows are co-added for each region and stored as 1D arrays listing the sum for each column of the CCD.
An example for the case of HR~8500 is shown in Fig.~\ref{fig:HR8500_ffi}. For 70~Aqr and HD~178875, we took the sum of the virtual and masked flux as our raw smear observable. For a target star that falls near the upper or lower edge of the detector, one of these smear arrays can be affected by saturated flux bleeding along the column. As a result, for HR~8500, only the masked flux was suitable, as discussed in Section~\ref{hr8500}. 

Two factors restrict the usefulness of the smear data to the brightest stars. One is the fact that the data are summed across all 1024 rows of the detector, so that targets in the same column are confused. The other is that, as noted by \citet{orig_smear}, the effective exposure time for smear data is much less than for ordinary target pixels.  This is because we have 12 rows each of masked and virtual smear, which measure the incident flux during the 0.52 second readout, so that the effective exposure time is only $12/1070 * 0.52/6.02 \sim 1/1034$ of the true pixel exposures for each of the virtual and masked smear, or $\sim 1/517$ if both can be used.
Hence, the photometric precision for bright stars is expected to be equivalent to that of a star $\sim 6.7$ mags fainter observed in the standard way, but for stars fainter than a Kepler magnitude (Kp) of $\sim 9$ (as defined in \citet{2010ApJ...713L..79K}), read noise dominates and the smear precision falls off faster than the square root of the number of photons.

For a more detailed description of the smear data we recommend to the reader \citet{2010ApJ...713L..92C}, \citet{doi:10.1117/12.857678}, the \emph{Kepler} Data Processing Handbook (KSCI-19081-001) \citep{fanelli}, and the \emph{Kepler} Instrument Handbook (KSCI-19033) \citep{vancleve}.

For each target star, we visually inspected the full-frame images (FFIs) taken at the beginning of each Quarter (\emph{Kepler}) (hereafter~Q) or Campaign (K2) and located the relevant columns in the corresponding smear data. An example for K2~C3 is shown in Fig.~\ref{fig:HR8500_ffi}. We see that the columns containing the bright saturated star HR~8500 (Section~\ref{hr8500}) correspond to a very strong peak in the smear data. Having identified the target in the smear data, we selected an appropriate range of columns, taking care to avoid confusion with nearby sources, and extracted a time series for each column. In the case of HD~178875 (Section~\ref{hd178875}) in the original \emph{Kepler} field, we took the sum of these time series as the light curve for the star. 

For stars observed in K2, pointing corrections introduce potentially large systematics. The photometric precision of K2 is lower than in the nominal Mission because the roll-axis equilibrium is unstable and must be corrected by thruster firings every $\sim 6$ hours, with momentum dumps on the other two axes every 2~days. We corrected for these by tracking the 1D~centroid of the smear peak, and extracting photometry using an aperture with a 3-pixel flat top and 2-pixel cosine bell (Hanning) taper on either side of this. We then applied the method of \citet{2015MNRAS.447.2880A}, using a Gaussian process (GP) systematic model, using a 1D~squared exponential kernel for continuum variations as a function of time. We model the pointing variations with a 2D~squared exponential kernel taking as inputs the predicted $x$ and $y$ position of a nearby star: for HR~8500, this was EPIC~206249807, and for 70~Aqr it was EPIC~206164235.

While in Sections~\ref{hr8500} and~\ref{70aqr} we present an analysis of systematics-corrected lightcurves, the corrections are minor and the raw lightcurves support the same qualitative conclusions. Other detrending methods such as those of \citet{2014PASP..126..948V}, \citet{2015ApJ...806...30L} and \citet{2015ApJ...806..215F} may also achieve similar results.

\section{Results}
\label{targets}

\subsection{HD~178875}
\label{hd178875}

\begin{figure*}
\begin{center}
	\includegraphics[width=0.85\textwidth]{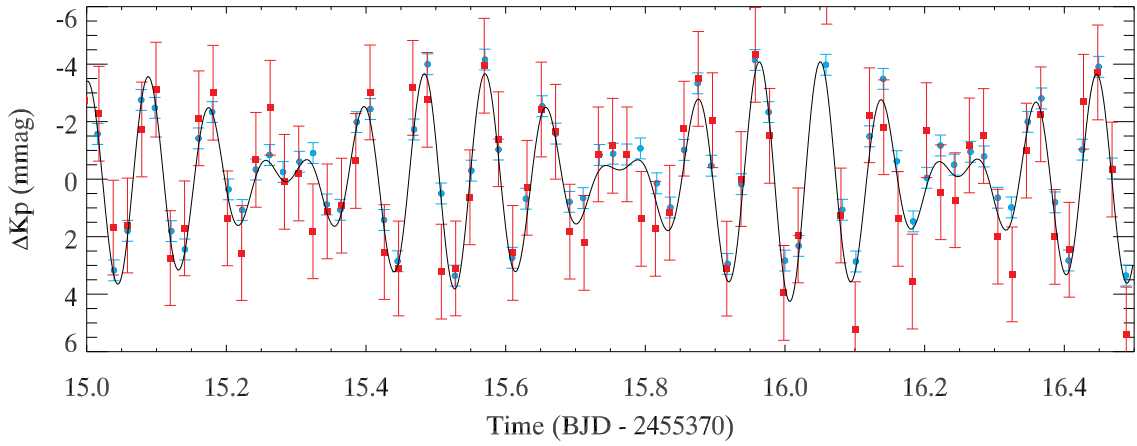}
	\caption{Light-curve segment from Q6 for HD\,178875 from the msMAP data (blue circles) and the smear data (red squares). The black line shows a six-frequency fit.}
	\label{fig:3429637_light_curve}
\end{center}
\end{figure*}

HD\,178875 (KIC\,3429637) is one of the brightest $\delta$\,Scuti stars in the original \textit{\emph{Kepler}} field, with a Kp~mag of 7.711. Over the first 2\,yr of the \textit{\emph{Kepler}} Mission it showed a growth in pulsation amplitude that was attributed to its advanced evolutionary state, near the terminal-age main sequence \citep{murphyetal2012}.

We have extracted the six strongest oscillation frequencies present in the smear data of this star, and they agree with those published by \citet{murphyetal2012} to within the 1-$\sigma$ uncertainties. A light-curve segment with a six-frequency fit is shown in Fig.\,\ref{fig:3429637_light_curve}. The msMAP \citep{2014PASP..126..100S} and smear light curves resemble each other closely, although the latter has a higher noise level because of the much lower effective exposure time. 
The error bars drawn in Fig.\,\ref{fig:3429637_light_curve} are proportional to the per-point residuals after the non-linear least-squares fit.

Stellar oscillation frequencies are the fundamental data of asteroseismology. Their recovery to 1-$\sigma$ precision in smear data against published values that used aperture photometry is a good demonstration of the utility of these data.

\subsection{70~Aqr}
\label{70aqr}

\begin{figure}
\begin{center}
\resizebox{\hsize}{!}{\includegraphics{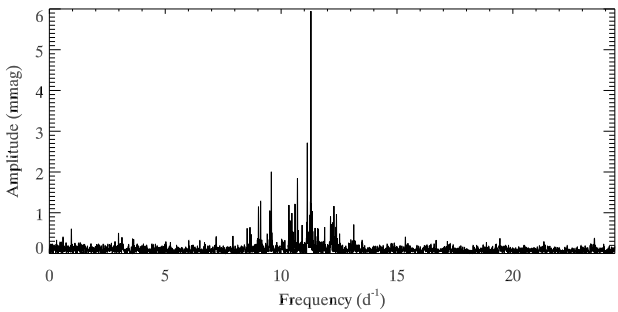}}
	\caption{Amplitude spectrum of the K2 C3 smear data for the $\delta$\,Scuti star 70\,Aqr.}
	\label{fig:70aqr_FT}
\end{center}
\end{figure}

Some stars are so bright on the \textit{\emph{Kepler}} CCD that it is impractical to assign sufficient pixels to perform aperture photometry. An example in the K2 field for C3 is 70\,Aqr (HR\,8676; HD\,215874), which has a magnitude of $V=6.2$. It is in some ways a typical $\delta$\,Scuti star -- its effective temperature (7300\,K; \citealt{paunzenetal2002a}) puts it in the middle of the $\delta$\,Scuti instability strip, its rotational velocity (100\,km\,s$^{-1}$; \textit{ibid.}) lies at the mode of the distribution for stars of its spectral type \citep{royeretal2007}, and the Fourier transform of its \textit{\emph{Kepler}} smear light curve (Fig.\,\ref{fig:70aqr_FT}) is typical in that it contains a few tens of statistically significant peaks. However, its Fourier transform is unusual in other ways -- its highest oscillation peak puts it in the top percentile of \textit{\emph{Kepler}} $\delta$\,Scuti stars by peak amplitude (see, e.g. \citealt{murphy2014}), and the frequency distribution is rather narrow.

The \textit{K2} photometry provided here offer another epoch of observations for this star. Literature data suggest 70\,Aqr undergoes amplitude variability, as is common in $\delta$\,Scuti stars (see \citealt{bowman&kurtz2014} for a recent discussion). For instance, the catalogue of $\delta$\,Scuti stars by \citet{rodriguezetal2000} lists an amplitude of 20\,mmag in V, whereas the \textit{\emph{Kepler}} smear data for the two peaks at 11.2859 and 11.1298\,d$^{-1}$ have amplitudes of 6.1 and 2.9\,mmag, respectively. The older literature has amplitudes ranging from 8.5 to 25\,mmag, each for different oscillation periods \citep{weiss1977,hildebrandt1992}, suggesting the dominant mode is not always the same. In all cases, these ground-based observations cannot match the excellent duty cycle of \textit{\emph{Kepler}}.

One of the greatest problems in the study of $\delta$\,Scuti stars is mode identification. This is usually performed with multi-colour photometry or time-series spectroscopy. \emph{Kepler}/K2 observes in white light only, and so ground-based observations are still very valuable for mode identification. Although 70\,Aqr is unremarkable as a $\delta$\,Scuti star, it is bright enough for successful ground-based observing campaigns, and now thanks to the smear data we can produce a high-precision light curve despite the gross saturation on the \emph{Kepler} CCD.

\subsection{HR~8500}
\label{hr8500}

\begin{figure}
\begin{center}
\resizebox{\hsize}{!}{\includegraphics{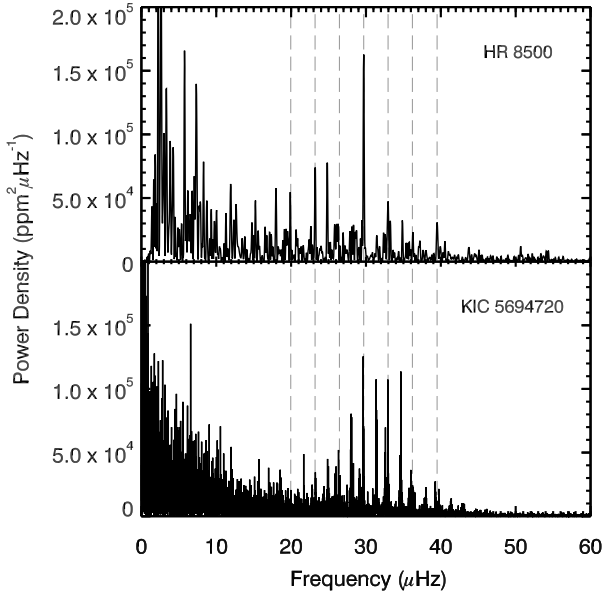}}
\caption{Top panel: Power spectrum of HR 8500 K2 C3 smear data after outlier rejection and high-pass filtering with a boxcar width of 10 days. Dashed lines indicate the locations of radial oscillation modes. Bottom panel: Power spectrum of the red giant star KIC~5694720, with frequencies scaled to match those measured in HR~8500 (the scaling factor was 0.98).}
\label{fig:HR8500_PS}
\end{center}
\end{figure}

Red giants exhibit radial and non-radial oscillations that are stochastically excited and damped by convection \citep{deridder09}. The amplitudes of these oscillations are significantly smaller than those in $\delta$~Scuti stars \citep{kb95}, so detection of red giant oscillations provides a more stringent test of the precision of smear data photometry.

HR~8500 (EPIC~206246606) is one the brightest red giants within the K2~C3 field of view, with a visual magnitude of 5.8. As for 70~Aqr, HR~8500 was not targeted due to the large number of pixels that would be required. 
As HR~8500 lies near the edge of the detector, its bleed column intersects with the virtual smear rows, so that the virtual smear data are affected strongly by the pointing cycle. Where the charge piles up like this, the measured smear flux is not expected to be linear with the incident flux. The associated systematics are, in the case of HR~8500's virtual smear column, of the same order as the raw flux and are not well-corrected with Gaussian process systematics models.  Under these circumstances, it is preferable to use only the masked smear array. In doing so, we lose a factor of two in flux, but have a substantially reduced contribution from pointing systematics.

Using $B-V=1.165$ from the EPIC catalog \citep{epic} with the colour-\teff\ relations by \citep{ramirez05} we estimate $\teff=4570$\,K, which combined with the Hipparcos parallax \citep{vanleeuwen07b} and a bolometric correction $BC_{V}=-0.45$ \citep{alonso99} yields a radius of $\sim 13.6\rsun$. Assuming a typical mass range for red giants of $1-2\msun$, we expect HR\,8500 to oscillate with a frequency of maximum power $\numax \sim 20-40\muHz$.

\begin{figure}
\begin{center}
\resizebox{\hsize}{!}{\includegraphics{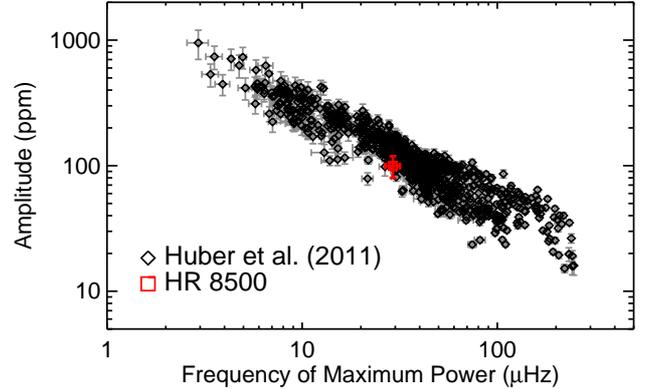}}
\caption{Oscillation amplitude versus frequency of maximum power for over 1000 red giants observed by the \emph{Kepler} Mission taken from \citep{huber11b}. The red square shows the measured values for HR\,8500 from the K2 C3 smear data.}
\label{fig:HR8500_amps}
\end{center}
\end{figure}

The top panel of Fig.~\ref{fig:HR8500_PS} shows the power spectrum of HR\,8500 obtained from the smear data. The spectrum shows a power excess around $\sim 30\muHz$, consistent with the expected $\numax$ value from colours and the Hipparcos parallax. Using the method of \citet{huber09} we measured $\numax=29.2\pm2.5\muHz$ and an amplitude per radial mode of $A=99\pm20\,$ppm. To further test whether the detection is compatible with solar-like oscillations, we compared our measurements to over 1000 red giants observed in the original \emph{Kepler} Mission, which follow a well-established relation between \numax\ and oscillation amplitudes \citep{hekker09,baudin11,stello11}. As shown in Fig.~\ref{fig:HR8500_amps} the amplitude measured in HR\,8500 is fully consistent with the \emph{Kepler} sample.

For comparison, the power spectrum of a similar red giant, observed throughout the {\it \emph{Kepler} Mission}, is shown in the bottom panel of Fig.~\ref{fig:HR8500_PS}. The frequency axis has been scaled by a factor of 0.98 in order to align peaks in both power spectra. From this we are able to identify the radial oscillation modes in HR~8500, and determine the characteristic frequency spacing, $\Delta\nu$, between modes of consecutive radial orders to be $3.25\pm0.02$\,$\mu$Hz.

Oscillating red giants are plentiful in the {\it \emph{Kepler}} and K2 fields \citep[e.g.][]{stello13}. Through the use of asteroseismic scaling relations \citep{ulrich86,brown91,kb95}, stellar properties including mass and radius can be determined from $\Delta\nu$ and $\nu_\mathrm{max}$, which is invaluable for population studies of the Milky Way galaxy \citep{miglio13}. However, these scaling relations require calibration to avoid systematic biases. Bright stars such as HR\,8500 may be well characterised through complementary methods such as long-baseline optical interferometry, providing the means to test and calibrate these relations \citep[e.g.][]{huber12,2013MNRAS.433.1262W}.

\section{Conclusions}
\label{conclusions}

As we have shown with the examples above, collateral smear data dramatically expand the possibilities for bright star science with \emph{Kepler} and K2. These ecliptic targets will not be observed by TESS, or will only be observed briefly, and this is therefore the only opportunity in the immediate future for space-based photometry of these bright stars.

The collateral archive already contains observations of many stars which can now be revisited. There are 29~stars with Kp$ < 7\,\mathrm{mag}$ that were not observed by the nominal \emph{Kepler} Mission, and a further 18~stars that were only observed in a few quarters. There are 17~stars in each of K2~C3 and C4 brighter than Kp $= 7\,\mathrm{mag}$ that were not targeted. Furthermore, collateral data for s 0,~1,~and~2 have not yet been made available, but there are 142 unobserved targets with Kp$ < 7\,\mathrm{mag}$. Given the encouraging results of the analysis of C3 targets presented here, we believe they will also present valuable opportunities for advancing stellar astrophysics.

In allocating active pixels for future K2 campaigns, it will be valuable to consider allocating short cadence pixels to columns containing targets of interest, as the smear data for these columns will then be available at short-cadence time resolution, a less-bandwidth-intensive method to obtain information about these stars on short time scales. We also note that custom apertures for bright stars, which are expensive in terms of pixels, may be unnecessary in many cases, easing the overall competition for bandwidth and permitting future Campaigns to observe a larger number of faint targets.

\section*{Acknowledgements}

This research made use of NASA's Astrophysics Data System; the SIMBAD database, operated at CDS, Strasbourg, France; the IPython package \citep{PER-GRA:2007}; SciPy \citep{jones_scipy_2001}. Some of the data presented in this paper were obtained from the Mikulski Archive for Space Telescopes (MAST). STScI is operated by the Association of Universities for Research in Astronomy, Inc., under NASA contract NAS5-26555. Support for MAST for non-HST data is provided by the NASA Office of Space Science via grant NNX13AC07G and by other grants and contracts. We acknowledge the support of the Group of Eight universities and the German Academic Exchange Service through the Go8 Australia-Germany Joint Research Co-operation Scheme. D.H. acknowledges support by the Australian Research Council's Discovery Projects funding scheme (project number DE140101364) and support by the NASA Grant NNX14AB92G issued through the \emph{Kepler} Participating Scientist Program.




\bibliographystyle{mnras}
\bibliography{ms} 





\bsp	
\label{lastpage}
\end{document}